\documentclass[english,aps,prb,twocolumn,amsfonts,showpacs]{revtex4}
\usepackage[T1]{fontenc}
\usepackage[latin9]{inputenc}
\usepackage{amstext}
\usepackage{graphicx}
\usepackage{amsmath}
\usepackage{amssymb}
\usepackage{esint}

\usepackage{natbib}

\makeatletter
\@ifundefined{textcolor}{}
{%
 \definecolor{BLACK}{gray}{0}
 \definecolor{WHITE}{gray}{1}
 \definecolor{RED}{rgb}{1,0,0}
 \definecolor{GREEN}{rgb}{0,1,0}
 \definecolor{BLUE}{rgb}{0,0,1}
 \definecolor{CYAN}{cmyk}{1,0,0,0}
 \definecolor{MAGENTA}{cmyk}{0,1,0,0}
 \definecolor{YELLOW}{cmyk}{0,0,1,0}
 }

\makeatother

\usepackage{babel}

\begin{document}

\title{Decoherence and Quantum Interference assisted electron trapping in a quantum dot}

\author{Ahmed El Halawany$^{1,2}$ and Michael N. Leuenberger$^{1,2}$}

\email{Michael.Leuenberger@ucf.edu}

\affiliation{$^{1}$NanoScience Technology Center, University of Central Florida,
Orlando, Florida 32826, USA}

\affiliation{$^{2}$Department of Physics, University of Central Florida, Orlando,
Florida 32816, USA}
\begin{abstract}
We present a theoretical model for the dynamics of an electron that gets trapped by means of decoherence and quantum interference in the central quantum dot (QD) of  a semiconductor nanoring (NR) made of five QDs, between 100 K and 300 K. The electron's dynamics is described by a master equation with a Hamiltonian based on the tight-binding model, taking into account electron-LO phonon interaction (ELOPI). Based on this configuration, the probability to trap an electron with no decoherence is almost 27\%. In contrast, the probability to trap an electron with decoherence is 70\%  at 100 K, 63\%  at 200 K and 58\%  at 300 K. Our model provides a novel method of trapping an electron at room temperature.
\end{abstract}


\maketitle

\section{Introduction}

The interaction between a quantum system and its environment is inevitable, leading to decoherence,\cite{Breuer:2002} which is one of the main obstacles in  fields such as quantum information processing,\cite{Fischer:2009} quantum optics, when measuring optical Schr\"odinger cat states,\cite{Deleglise:2008} 
condensed matter physics, when looking for mesoscopic interference phenomena in quantum transport of electrons,\cite{Oudenaarden:1998,Bachtold:1999} etc. Since many interesting quantum phenomena are based on coherence, many solutions are proposed, and are currently in use, to suppress or overcome decoherence,\cite{Anglin:1997} such as quantum error-correction codes,\cite{Shor:1995} error-avoiding codes,\cite{Shor:1995} echo techniques,\cite{Bluhm:2011,Petta:2005} quantum feedback operations,\cite{Deleglise:2008} optimal control technique,\cite{Rabitz:2000} and many more. Other research groups are trying to fight decoherence through the knowledge of their spectral density, thinking this would be more operative.\cite{Alvarez:2011} A rather opposite approach to this stream of research is found in quantum biology, where scientists are trying to take advantage of the decoherence in the quantum dynamics of excitons in order to find explanations for the high efficiency in solar energy harvesting in photosynthetic systems.\cite{Mohseni:2008,Caruso:2009} Recent explanations include environment-assisted energy transfer in quantum networks, such as noise-assisted transport \cite{Mohseni:2008,Plenio:2008} and oscillation-enhanced transport.\cite{Semiao:2010,Lloyd:2010}

The role of decoherence in localizing electrons has been reported in many previous works.\cite{Zeh:1996,Dieter:2011} Another approach is to apply continuous measurement to keep the quantum state in a pure state. This approach is known as the quantum Zeno effect.\cite{Sudarshan:1977} Some groups report that continuous measurement will lead to quantum anti-Zeno effect.\cite{Gontis:1997} In all previous reports, quantum interference does not play any role in trapping the electron. In this work, we present a configuration that focuses on the interplay between quantum interference of the electron wavefunction and decoherence in trapping the electron in the central QD.
Neither quantum interference nor decoherence alone can trap the electron wavefunction in one out of five coupled quantum dots. It is the combined effect of quantum interference and decoherence that leads to the trapping.

\begin{figure}[htb]
\includegraphics[width=8.5cm]{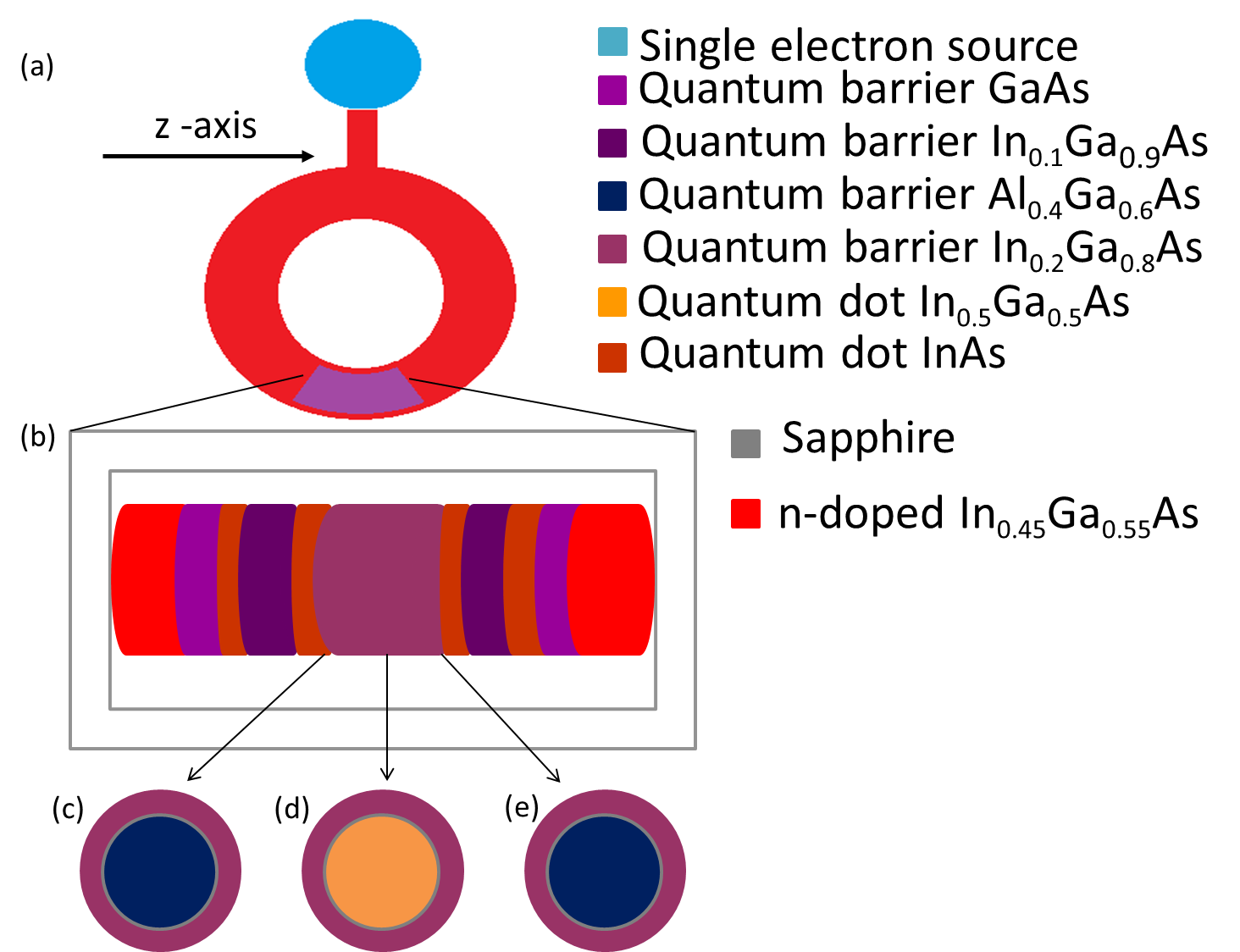}

\caption{(a) Schematic setup (not to scale). (b) A magnified diagram for the intrinsic region made of five quantum dots of (from left to right) 1.24, 1.5, 5.0, 1.5, 3.14 nm height, respectively. (c) A cross section view for the interface between quantum dot \#2 and barrier \#3. (d) A cross section view for quantum dot \#3 (electron pocket). (e) A cross section view for the interface between quantum dot \#2 and barrier \#4. \label{fig:setup}}

\label{fig:NRsetup}
\end{figure}

\section{Structure and mechanism}

\begin{figure}[htb]
\includegraphics[width=8.5cm]{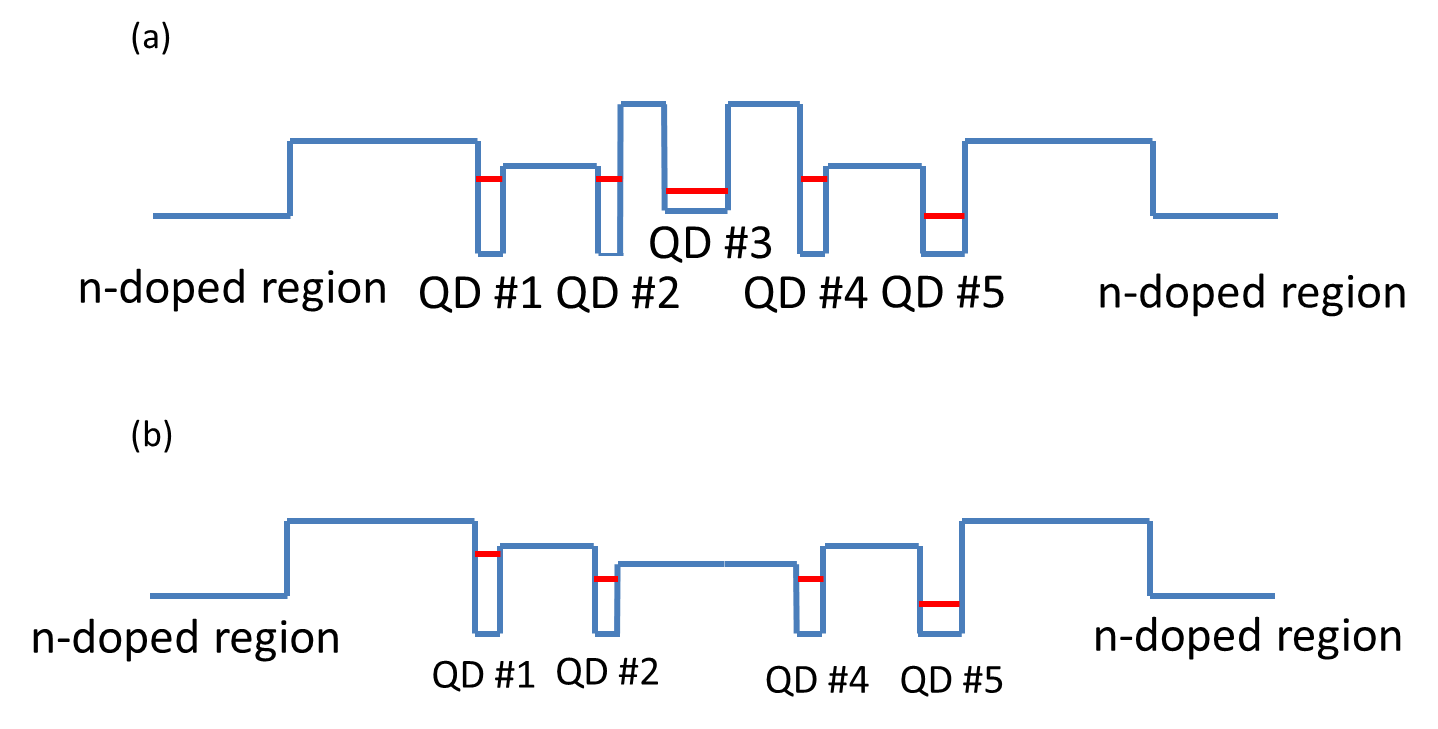}

\caption{(a) The conduction band, for radius $r< 8.05$ nm. (b) The conduction band, for $r > 8.05$ nm. The quantum dots' ground eigenstates are shown. \label{fig:CB_configuration}}

\end{figure}

We consider the transport of a single electron in a NR with 15.1 nm as minor radius and  30 nm as major radius (see Fig.~\ref{fig:NRsetup}). The NR is divided into two regions. 
The first region, which is n-doped In$_{0.45}$Ga$_{0.55}$As with a concentration of $6.0\times 10^{14}$ cm$^{-3}$, constitutes 85\% of the NR, and it will be referred to as the "zero-region" in the manuscript. The second region is called the "intrinsic-region" in the manuscript. It consists of five QDs, four of which are made of InAs, and the central QD is made of In$_{0.5}$Ga$_{0.5}$As. QD \#5, (see Fig.~\ref{fig:CB_configuration} for QDs labeling), is n-doped with a concentration of $1.0\times 10^{18}$ cm$^{-3}$. The zero-region and QD \#5 are not degenerate semiconductors. Barrier \#1 and \#6 are made of GaAs, while barriers \#2 and \#5 are made of In$_{0.1}$Ga$_{0.9}$As. As for barrier \#3 and \#4, they are made of Al$_{0.4}$Ga$_{0.6}$As. A monolayer of sapphire Al$_2$O$_3$, which has radius of 8.05 nm, coats the region starting from the interface between QD \#2 and barrier \#3 to the interface of barrier \#4 and QD \#4. The outer layer, up to the surface of the NR, is made of In$_{0.2}$Ga$_{0.8}$As. As a result of this concentric configuration, the central QD acts like an \textit{electron pocket} that traps the electron with the help of decoherence, as will be shown later. All interfaces between the materials considered in the aforementioned configuration are recognized to be straddling gaps (type I). Based on all chosen materials and types of interfaces, the conduction band (CB) profile is shown in Fig.~\ref{fig:CB_configuration}, based on the self-consistent solution of the Schr\"odinger-Poisson equation. All semiconductor materials have the same crystal structure and direct band gap. In addition, the NR is coupled to a single-electron source (SES).\cite{Feve:2007,Bocquillon:2012}
The SES is triggered to emit an electron and thus this electron can transport through the whole configuration. Therefore, the time evolution is well described by the single-electron master equation as shown and justified below.

\section{Model}

Given that the aforementioned configuration has zero electric field across the five-QD region and the electron's eigenenergies are close to the conduction band-edge minima, the 3D time-independent Schr\"odinger equation in cylindrical coordinates and in the effective mass approximation is used to find the eigenenergies and eigenstates for each QD separately. These states are used later (see below) to describe the dynamics of the electron by means of a generalized master equation in the tight-binding approximation, taking into account ELOPI. For simplicity, an infinite confining potential in the radial direction is assumed. The eigenenergies and wavefunctions of QDs \#1 and \#5 are obtained systematically. As for QD \#3, due to the relatively large band gap (5 eV) for the sapphire Al$_2$O$_3$ monolayer, it is assumed to be confined in infinite potential but with different radius than QDs \#1 and \#5. Both QDs \#2 and \#4 (see App. B), require an additional boundary condition due to the \textit{electron pocket} i.e. the electron's energy has to be conserved irrespective of the interface with Al$_{0.4}$Ga$_{0.6}$As or In$_{0.2}$Ga$_{0.8}$As (see Fig.~\ref{fig:NRsetup}).  The zero energy is set at the minimum of the conduction band of InAs QDs.
 
We start with the following Hamiltonian \begin{equation}
H = H_I + H_c,
\end{equation}
where $H_I$ is the Hamiltonian of an electron in the "intrinsic-region"
\begin{eqnarray} 
H_I & = & \mathop \sum \limits_i {\varepsilon _i}a_i^\dagger {a_i} + \left( { - \mathop \sum \limits_{i\ne j} {t_{ij}}a_i^\dagger {a_j} + {\rm{h}}.{\rm{c}}} \right)  \nonumber \\
& & + {\hbar\omega _{LO}}{b^\dagger}b+\lambda \mathop \sum \limits_i a_i^\dagger {a_i}\left( {{b^\dagger} + b} \right),
\label{eq:Hamiltonian}
\end{eqnarray}
and $H_c$ is the Hamiltonian that describes the coupling between both QD \#1 and \#5 and "zero-region"
\begin{equation}
H_c=\left(\sum \limits_0 V_{01} {C_0^\dagger} a_1 + {\rm{h}}.{\rm{c}} \right) + \left(\sum \limits_0 V_{05} {C_0^\dagger}a_5 + {\rm{h}}.{\rm{c}} \right).
\end{equation}
In Eq.~(\ref{eq:Hamiltonian}), the first term describes the on-site ground state for the five QDs. The second term, which is based on the tight-binding model, describes the hopping of the electron between the QDs, where ${{t}_{ij}}$ is a 3-D hopping integral given by the off-diagonal matrix elements of $H_t$,\cite{Omar:1975} i.e. (see Sec.~\ref{sec:hopping} for details)
\begin{equation}
t_{ij} = \mathop \smallint \Psi_i^* {H_t} {\Psi_j}d^3{r},
\end{equation}
where $H_t$ is the kinetic and potential energy of the electron inside the QD,
\begin{eqnarray}
H_t = -\frac{\hbar^2 }{2m^*} {\nabla_{3D}^2} + V\left(r,z\right).
\end{eqnarray}
The third term in Eq.~(\ref{eq:Hamiltonian}) describes non-dispersive LO phonons of In$_{0.45}$Ga$_{0.55}$As, since it constitutes 85\% of the NR.  In polar semiconductors, as the size of the QD decreases, electrons interact mostly with phonons that have long wavelength $\left|\mathbf{q}\right|\le 2\pi$ /(QD size). This suggests that a model with dispersionless LO phonons is accurate.\cite{Inoshita:1997} It has been shown in experimental work that for InAs QDs embedded in GaAs matrix, the GaAs LO phonons are more prominent than the InAs LO phonons. \cite{Bimberg:2011}  The fourth term in Eq.~(\ref{eq:Hamiltonian}) describes the interaction between the electron and LO phonons with coupling strength $\lambda$ (see App. A). In this work, $ g = \lambda/(\hbar \omega_{LO})\approx 0.066$. As for the acoustic phonons, in polar semiconductor nanostrutures, the electron-acoustic phonon coupling is weak because the energy difference between the ground state and excited state $\Delta E$ is greater than 64 meV in all QDs. Moreover, in the central QD the energy difference is greater than 110 meV. As a result, the acoustic phonons are taken into account in the master equation, as shown later, as part of the total decoherence. Since we deal with ELOPI, a canonical unitary transformation is useful to eliminate the linear coupling terms in Eq.~(\ref{eq:Hamiltonian}). The transformed Hamiltonian is $H_{I}' = e^S{H_I}e^{-S}$, where $S = -g\left(\mathop \sum \limits_{i}a_i^\dagger{a_i}\left({b^\dagger} + b \right)\right)$. We obtain
\begin{eqnarray}
H_I^{'} & = & \mathop \sum \limits_i a_i^\dagger {a_i} \left(\varepsilon_i - \lambda^2/\hbar \omega_{LO} \right) \nonumber \\
& & +\left(- \mathop \sum_{ij} {t_{ij}}a_i^\dagger {a_j} e^{-2g\left(b^\dagger  - b\right)}\right)
\nonumber \\
& & +\left(-\mathop \sum_{ij} {t_{ji}}a_j^\dagger {a_i} e^{2g\left(b^\dagger - b\right)}\right) 
\nonumber \\
& & + \hbar \omega_{LO} b^\dagger {b} + \left(2 \lambda^2 / \hbar \omega_{LO} \right) .
\label{eq:Hamiltonian_transformed}
\end{eqnarray}
In Eq.~(\ref{eq:Hamiltonian_transformed}) the first term shows the renormalization of the QDs' eigenstates in the presence of  ELOPI.
% In other words, the eigenenergies of the QDs are shifted by $\lambda^2/\hbar \omega_{LO}$.  
The eigenstates of the transformed Hamiltonian $H_I'$ are in the tensor product form and are denoted by $\left|\nu_{g},N\right\rangle$ and $\left|\nu_{e},N\right\rangle$, where $\nu_g$ and $\nu_e$  represent the ground and excited state of the electron in QD \#$\nu$, respectively, and $N$ represents the number of LO phonons.  The Hamiltonian in Eq.~(\ref{eq:Hamiltonian_transformed}) is solved in the following basis $\left|SES\right\rangle$, $\left|zero-region\right\rangle$, $\left|1_{g},0\right\rangle$, $\left|2_{g},0\right\rangle$, $\left|3_{g},0\right\rangle$, $\left|4_{g},0\right\rangle$ and $\left|5_{g},0\right\rangle$, where $\left|SES\right\rangle$ is the electron in the SES before being injected in the NR, $\left|zero-region\right\rangle$ is the electron in the "zero-region" in the NR after being injected from the SES, and $\left|1_{g},0\right\rangle$ is the electron in the ground state of QD \#1 with no phonons.
We define the phonon displacement operator $D(\beta)=e^{\beta b^\dagger - \beta^* b}$. We can now make use of the well-known formula for the matrix elements of the displacement operator\cite{Cahill&Glauber:1969,Cahill&Glauber:1969_2}
\begin{eqnarray}
\left\langle N'\right| D(\beta)\left|N\right\rangle & = & \left(\frac{N!}{N'!}\right)^{1/2}|\beta|^{N'-N}e^{-|\beta|^2/2} \nonumber\\
& & \times L_N^{N'-N}(|\beta|^2)e^{i(N'-N)\phi},
\end{eqnarray}
where $\beta=|\beta|e^{i\phi}$ and $ L_N^{N'-N}(|\beta|^2)$ are the associated Laguerre polynomials.
For $\beta\ll 1$ the associated Laguerre polynomials are approximately
\begin{equation}
L_N^{N'-N}(|\beta|^2)\approx \frac{N'!}{N!(N'-N)!}\left(1+\frac{N}{N'-N+1}|\beta|^2\right).
\end{equation}
Thus, for $\beta\ll 1$ only the phonon states with $N'=N$ couple to each other in a good approximation, and
$\left\langle N\right| D(2g)\left|N\right\rangle=e^{-2g^2}$.
Therefore, the second and third terms in Eq.~(\ref{eq:Hamiltonian_transformed}) show that the hopping term $t_{ij}$ is reduced by a factor of $e^{-2\lambda^2/(\hbar \omega_{LO})^2}$. In the weak ELOPI considered in this work, $e^{-2\lambda^2/(\hbar \omega_{LO})^2}\approx 1$ and thus the hopping terms are not reduced (see App. A). 

The validation of the aforementioned Hamiltonian depends on the following criteria; in this configuration there must be no electrons in the CB. This is calculated in the standard way as follows
\begin{equation}
n = \mathop \smallint \limits_{{E_c}}^\infty  D\left( E \right){f^{FD}}\left( E \right)dE.
\end{equation}

\begin{figure}[htb]
\includegraphics[width=8.5cm]{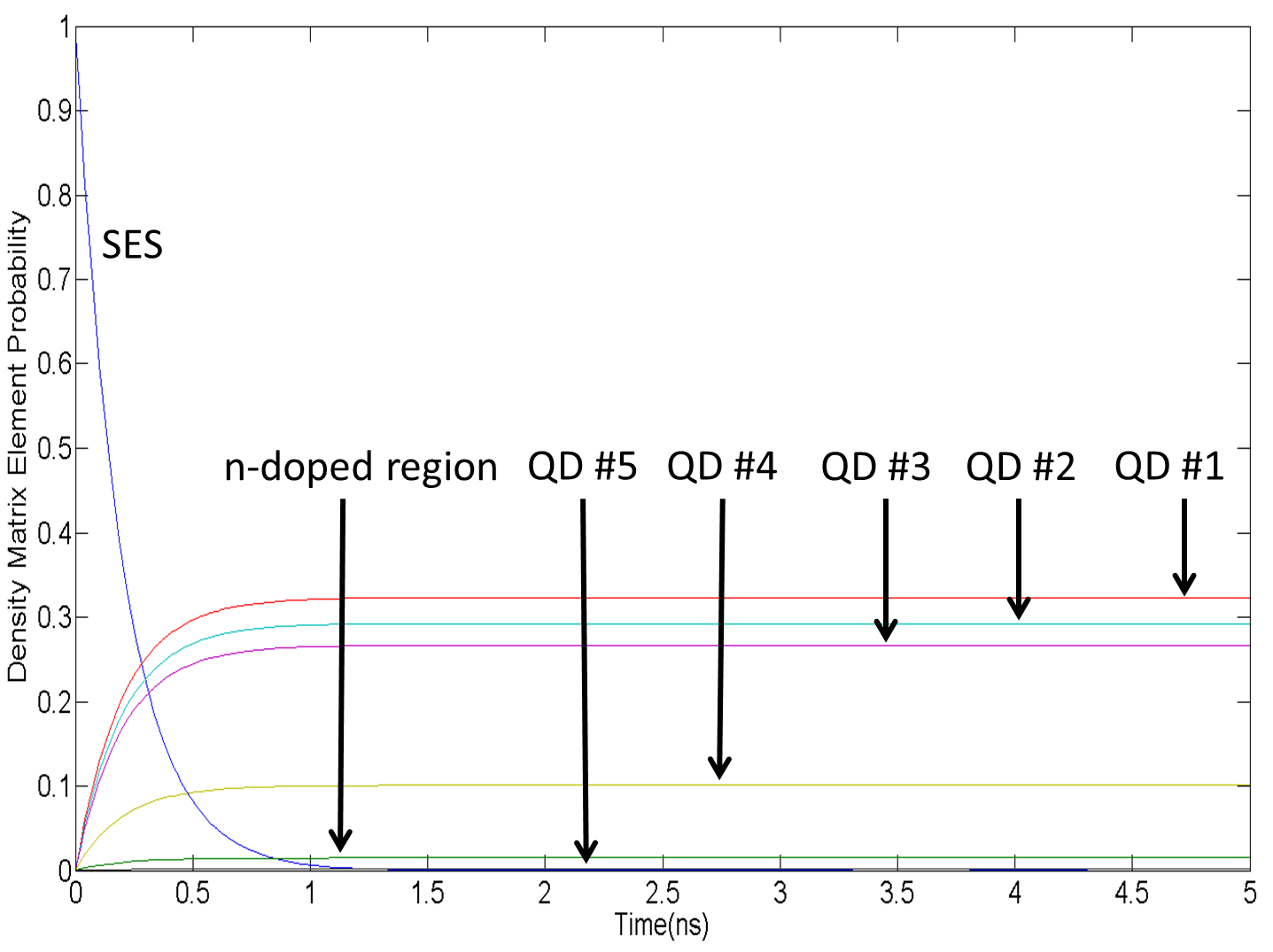}

\caption{The electron's time-dependent probability distribution among the five quantum dots in the case with zero decoherence. \label{fig:probabilities_no_decoherence} }

\end{figure}

Since the configuration has a large length-to-width ratio, $D\left( E \right)$ is approximated by the density of states of a 1D NR (Section V). Doping (type and concentration) along with temperature are taken into account through the Fermi-Dirac function ${{f}^{FD}}\left( E \right)$. We observe that on average there are virtually no electrons in the whole configuration. Consequently, when the SES is triggered and a single electron is emitted, $H_I'$ is a valid description for this single electron. $H_c$  describes the coupling between the "zero-region" (quasi-continuum) and QDs \#1 and \#5. The electron inside the "zero-region" is incoherent because of the decoherence effects due to the acoustic phonons. Such effects do not conserve the energy of the electron. This coupling is well described by Fermi's golden rule for the transition rates as follows: 
\begin{eqnarray}
{W_{0 \to 1_g}} & = & \frac{{2\pi}}{\hbar} \smallint \limits_{{E_c}}^\infty dE_0 {\left| {\left\langle {1_g{\rm{|}}H_c{\rm{|}}0} \right\rangle} \right|^2} D\left(E_0 \right) {f^{FD} \left(E_0 \right)} \nonumber\\
& & \times \left( 1 - {f^{FD}} \left(E_{1g} \right) \right) \delta \left( {{E_{1g}} - E_0} \right) ,
%{W_{i \to f}} & = & \frac{{2\pi }}{\hbar }\mathop \sum \limits_i{\left| {\left\langle {f{\rm{|}}H_1{\rm{|}}i} \right\rangle } \right|^2}{f^{FD}({E_i})\left\{ {1 - {f^{FD}}({E_f})} \right\} \nonumber\\
%& & \times \delta \left( {{E_f} - {E_i} - e\Delta V} \right)
\end{eqnarray}
\begin{eqnarray}
{W_{1_g \to 0}} & = & \frac{{2\pi}}{\hbar} \smallint \limits_{{E_c}}^\infty dE_0 {\left| {\left\langle {0{\rm{|}}H_c{\rm{|}}1_g} \right\rangle} \right|^2} D\left(E_0 \right) {f^{FD} \left(E_{1g} \right)} \nonumber\\
& & \times \left( 1 - {f^{FD}} \left(E_0 \right) \right) \delta \left( {{E_{1g}} - E_0} \right) ,
\end{eqnarray}
\begin{eqnarray}
{W_{5_g \to 0}} & = & \frac{{2\pi}}{\hbar} \smallint \limits_{{E_c}}^\infty dE_0 {\left| {\left\langle {0{\rm{|}}H_c{\rm{|}}{5_g}} \right\rangle} \right|^2} D\left(E_0 \right) {f^{FD} \left(E_{5g} \right)} \nonumber\\
& & \times \left( 1 - {f^{FD}} \left(E_{0} \right) \right) \delta \left( {{E_0} - E_{5g}} \right), 
\end{eqnarray}
and 
\begin{eqnarray}
{W_{0 \to 5_g}} & = & \frac{{2\pi}}{\hbar} \smallint \limits_{{E_c}}^\infty dE_0 {\left| {\left\langle {5_g{\rm{|}}H_c{\rm{|}}{0}} \right\rangle} \right|^2} D\left(E_{5_g} \right) {f^{FD} \left(E_0 \right)} \nonumber\\
& & \times \left( 1 - {f^{FD}} \left(E_{5g} \right) \right) \delta \left( {{E_0} - E_{5g}} \right) 
\end{eqnarray}
The coupling terms in Eq. (3), $V_n$ and $V_p$, are much smaller than ${{t}_{12}}$ and ${{t}_{45}}$. This confirms that we have a weak coupling between the outer QDs and the "zero-region". Thus a standard formalism appropriate for the description of such a system is the generalized master equation in the Born and Markov approximation \cite{Blum:1996}
\begin{equation}
{\partial _t}{\rho _{m,n}} = \frac{i}{\hbar }{\left[ {\rho ,H_I'} \right]_{m,n}} + {\delta _{m,n}}\mathop \sum \limits_{l \ne m} {\rho _n}{W_{m,l}} - {\gamma _{m,n}}{\rho _{m,n}},
\label{eq:Master}
\end{equation}
where ${{\gamma }_{m,n}}=~\frac{1}{2}\sum_l \,\left( {{W}_{l,n}}+{{W}_{l,m}} \right)+\frac{1}{{{T}_{2}}}$ is the total decoherence which includes the dephasing time ${{T}_{2}}$ due to electron-phonon (both acoustic and optical, and both elastic and inelastic) interaction and the rates ${{W}_{m,l}}$ of transition between the "zero-region" and the outer QDs. Eq.~(\ref{eq:Master}) is valid when the correlation time in the heat bath is much smaller than the relaxation time of the electron system. A rough estimate for the correlation time is $\frac{\hbar }{{{k}_{B}}T}\sim (1-3.5)\times {{10}^{-14}}$ s for T = 100 K - 300 K respectively, which is much smaller than the electron relaxation time, in such systems, $\sim {{10}^{-12}}$ s. The dephasing time ${{T}_{2}}$, based on temperature, is determined through the homogeneous broadening $2\hbar /{{T}_{2}}$.\cite{Borri:1999,Uskov:2000} At room temperature, the dephasing times are of the order of 200-300 fs.\cite{Borri:1999,Uskov:2000,Borri:2001} We choose $T_2=285$ fs at 300 K because there is no carrier-carrier interaction. At T = 100 K, the dephasing time is 2 ps.\cite{Borri:1999,Uskov:2000,Borri:2001} It is worth to mention that we ignore the change in band gap due to the lattice constant mismatch between the different materials. However, this does not affect the final results. For calculating the ground state of QD \#5 the doping is taken into account through the Schr\"odinger-Poisson equation. As a result, the ground state of QD \#5 will be $E_{5g}^{'}={{E}_{5g}}-\frac{{{\lambda }^{2}}}{\hbar {{\omega }_{LO}}}+\Delta $, where $\Delta $ is the increase in the ground energy of QD \#5 (few meV) due to doping. The change in wavefunction of QD \#5 is negligibly small. In this work only the ground state in each QD \#$\nu$, denoted by $\left| \nu_{g},0 \right\rangle $, is considered. Such contribution is attributed to the following reasons; first, the electron's transition from the "zero-region" to the ground state $\left| 1_{g},0 \right\rangle $  is 100 times faster than the transition to the excited state $\left| 1_{e},0 \right\rangle $. In addition, the transition to $\left| 5_{g},0 \right\rangle$ is 10 times less than the transition to $\left| 1_{g},0 \right\rangle $. As a result, the electron in the "zero-region" will basically favor tunneling toward QD \#1 more than QD\#5. Second, in systems where the energy separation is 44 meV the relaxation takes 20 (40) ps at 300 (100) K.\cite{Sauvage:2002} Thus, based on the detailed balance condition,  $\frac{{{W}_{nm}}}{{{W}_{mn}}}=e^{-\frac{\hbar {{\omega }_{nm}}}{{{k}_{B}}T}}$, phonon-assisted excitation will take much more time.  In addition, in the aforementioned configuration, based on the dimensions of the QDs, the energy separation is more than 64 meV. Third, in polar semiconductors, even at room temperature, the emission of LO phonon is more favorable than the absorption of LO phonon. As a result, the $\left| \nu_{g},0 \right\rangle $ states remain populated and the excited states can be neglected.

In this model, we calculate all ten hopping integrals (Section IV). Based on the \textit{electron pocket} configuration, ${{t}_{24}}$ is larger than ${{t}_{43}} + {{t}_{32}}$. This is impossible to achieve in a similar configuration without an \textit{electron pocket}.

\begin{figure}[htb]
\includegraphics[width=8.5cm]{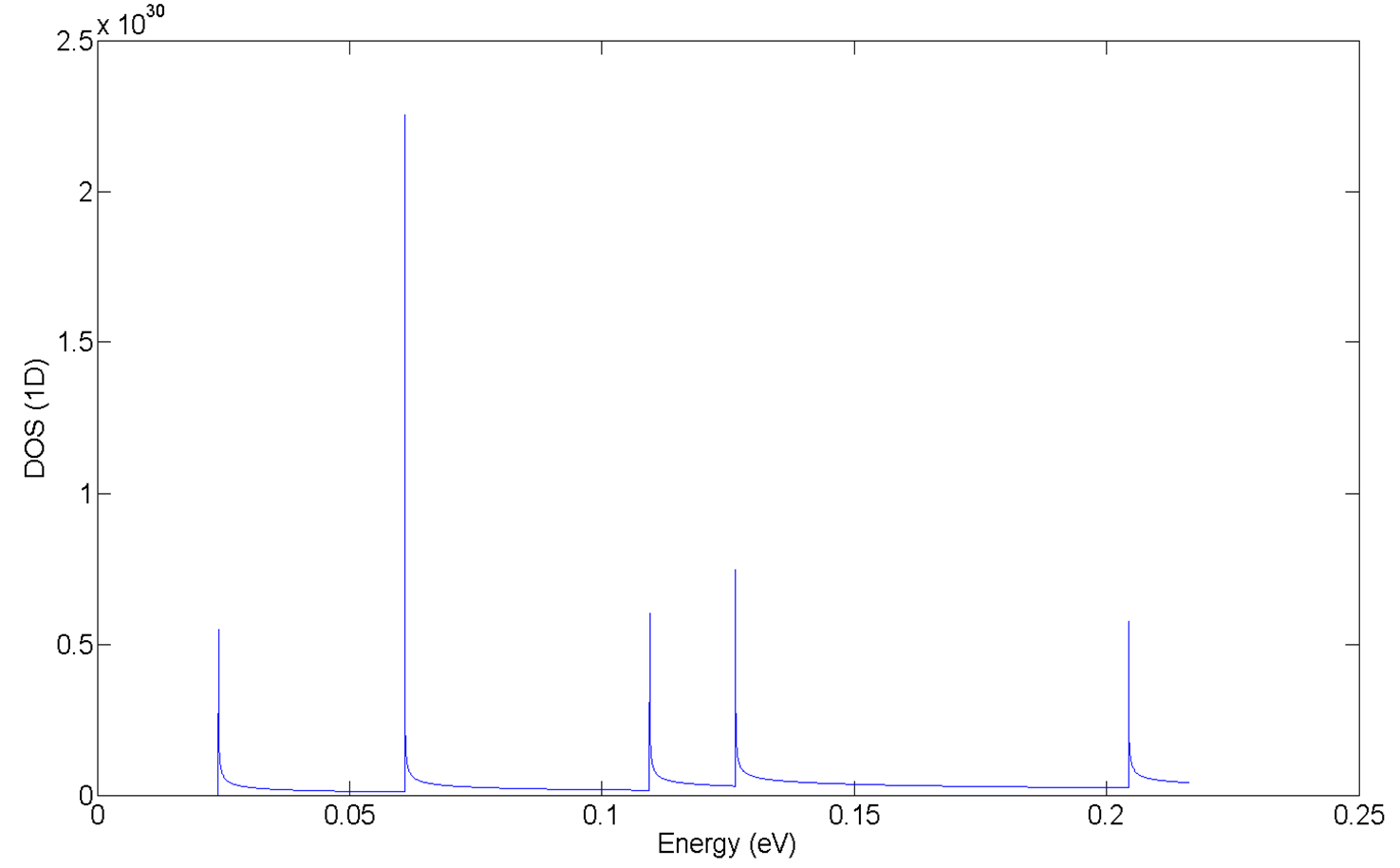}

\caption{Density of states for a circular nanowire with radius of 15.1 nm.}\label{fig:DOS}

\label{fig:DOS}
\end{figure}

\section{Calculation of hopping matrix elements}
\label{sec:hopping}

The total Hamiltonian of our system is
\begin{equation}
H_t = -\frac{\hbar^2 }{2m^*} {\nabla_{3D}^2} + V\left(r,z\right),
\end{equation}
where the first term is the kinetic energy and the second term is the potential energy
\begin{equation}
V\left(r,z\right)=\sum_\nu V_\nu.
\end{equation}
The potential $V_\nu$ represents the local potential of the QD \#$\nu$.
This representation can be used to derive the tight-binding Hamiltonian $H_I$ given in Eq.~(\ref{eq:Hamiltonian}) for the "intrinsic-region".
We provide an approximation to $H_I$ in App.~\ref{app:Landau}.
The diagonal and off-diagonal elements of $H_t$ are given by
\begin{equation}
\epsilon_i = -\int\Psi_i^*\frac{\hbar^2 }{2m^*} {\nabla_{3D}^2}\Psi_i d^3r + \int\Psi_i^*V_i\Psi_i d^3r
\end{equation}
and
\begin{equation}
t_{ij} = -\int\Psi_i^*\frac{\hbar^2 }{2m^*} {\nabla_{3D}^2}\Psi_j d^3r + \int\Psi_i^*V_i\Psi_j d^3r,
\label{eq:hopping}
\end{equation}
respectively. These are the variables that enter the tight-binding Hamiltonian $H_I$ in Eq.~(\ref{eq:Hamiltonian}).
Due to the cylindrical symmetry of the QDs we can write the wavefunction as
\begin{equation}
\Psi = R\left(\rho \right) \Phi\left(\phi \right) Z\left(z\right).
\end{equation}

Since the hopping is only along the axial axis (z-axis), $H_t$ is
\begin{equation}
H_{t} = -\frac{\hbar^2 }{2m^*} \partial_{zz} + V\left(r,z\right).
\end{equation}
The hopping integral is calculated in the following manner
\begin{equation}
t_{ij} = \int R_{i}^* \left(\rho \right)\Phi_{i}^*\left(\phi \right)Z_{i}^*\left(z\right){H_t}R_{j}\left(\rho \right)\Phi_{j}\left(\phi \right)Z_{j}\left(z\right)d^3{r},
\end{equation}
separating variables will yield
\begin{eqnarray}
t_{ij} & = & \int \limits_{{0}}^{2\pi} \Phi_{i}^*\left(\phi \right)\Phi_{j}\left(\phi \right) d\phi \int \limits_{{0}}^{r} R_{i}^*\left(\rho \right) R_{j}\left(\rho \right) d\rho \nonumber\\
& & \times \int \limits_{{-\infty}}^{\infty} Z_{i}^*\left(z\right) {H_t} Z_{j}\left(z\right) dz.
\end{eqnarray}
The azimuthal part will always result in 1. The azimuthal part acts like a selection rule for the electron hopping. The electron, in the considered configuration, hops between states with the same quantum number $m$. Consequently, the hopping occurs among the ground states of the QDs only. The radial integral requires special care, especially when one of the hopping integral's boundaries is located at the interface with the \textit{electron pocket},
\begin{equation}
\int \limits_{{0}}^{\gamma r} R_{iin}^*R_{jin} dr + \int \limits_{{\gamma r}}^{r} R_{iout}^*R_{jout} dr.
\end{equation} 
Since $H_I'$ is a hermitian, $t_{ij} = t_{ji}^*$. The values of the hopping integral vary based on the QDs. They are $t_{21}=t_{12}=161$ meV, $t_{54}=t_{45}=180$ meV, $t_{31}=t_{13}=14$ meV, $t_{53}=t_{35}=5$ meV, $t_{41}=t_{14}=24$ meV. $t_{25}=t_{52}=0.464$ meV, $t_{51}=t_{15}=0.1047$ meV, $t_{32}=t_{23}=56.44$ meV, $t_{34}=t_{43}=18.6$ meV, and $t_{42}=t_{24}=85.7$ meV.

\section{Density of states in n-doped region}

In this work, the NR's minor radius is 15.1 nm, and with circumference of almost 190 nm. The length-to-width ratio is almost 7 justifying the 1-D denisty of states (DOS) employed in this work. The DOS is given by\cite{Singh:2003}
\begin{eqnarray}
\rho^{1D} \left(E \right) & = & \mathop \sum_{i = 1}^n \left( \frac{2m^*}{\hbar^2}\right)^{1/2} \frac{1}{\pi \left(E - E_{i}\right)^{1/2}} \Theta \left( E-E_i \right). \nonumber\\
\end{eqnarray}
where $\Theta$ is a step function. The DOS graph is plotted in Fig.~\ref{fig:DOS}.

\begin{figure}[htb]
\includegraphics[width=8.5cm]{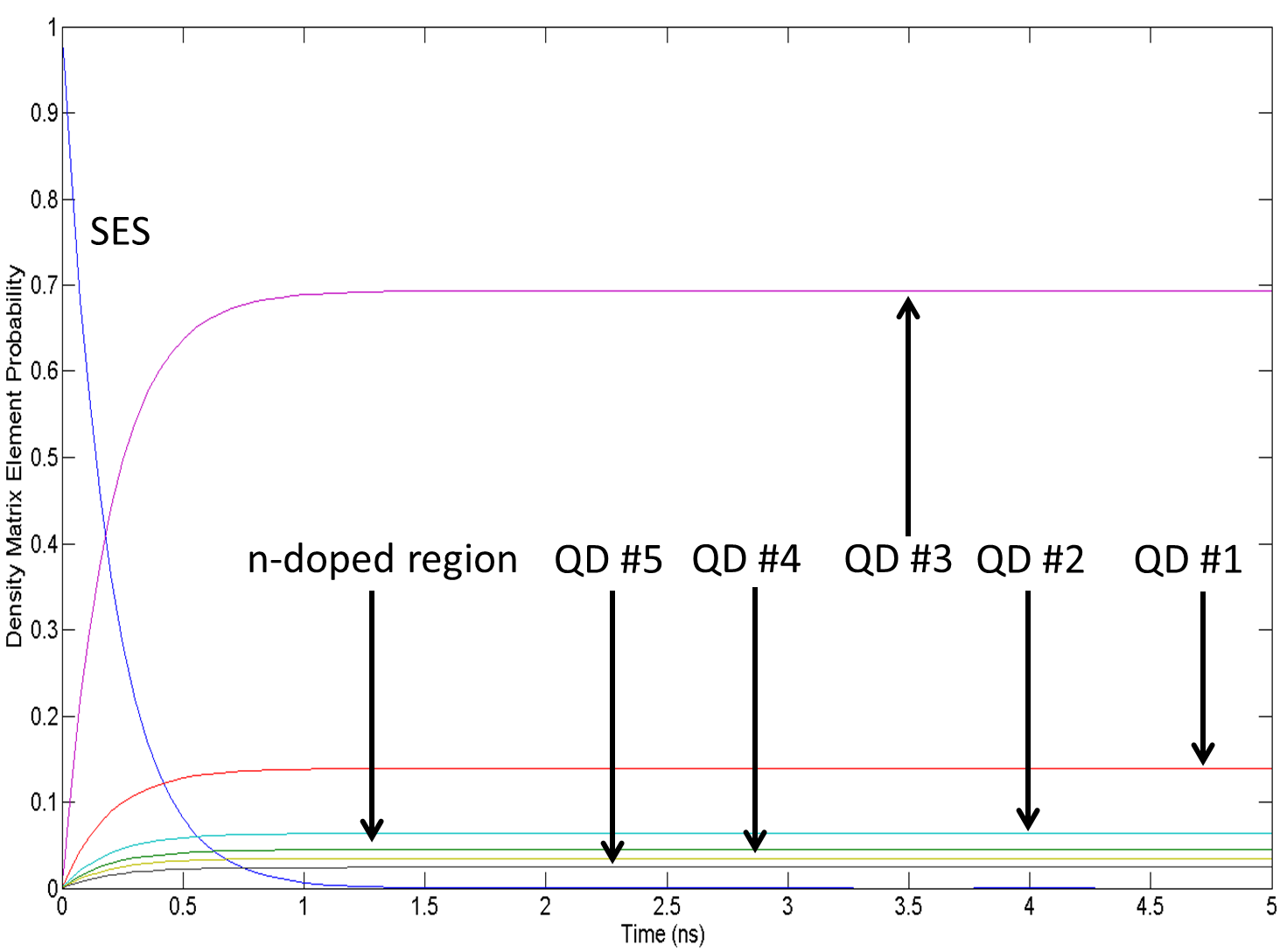}

\caption{The electron's time-dependent probability distribution among the five quantum dots at 100 K. The probability to trap the electron in QD \#3 is 70\%.
\label{fig:probabilities_decoherence_100K} }

\end{figure}

\section{Results and discussion}

\begin{figure}[htb]
\includegraphics[width=8.5cm]{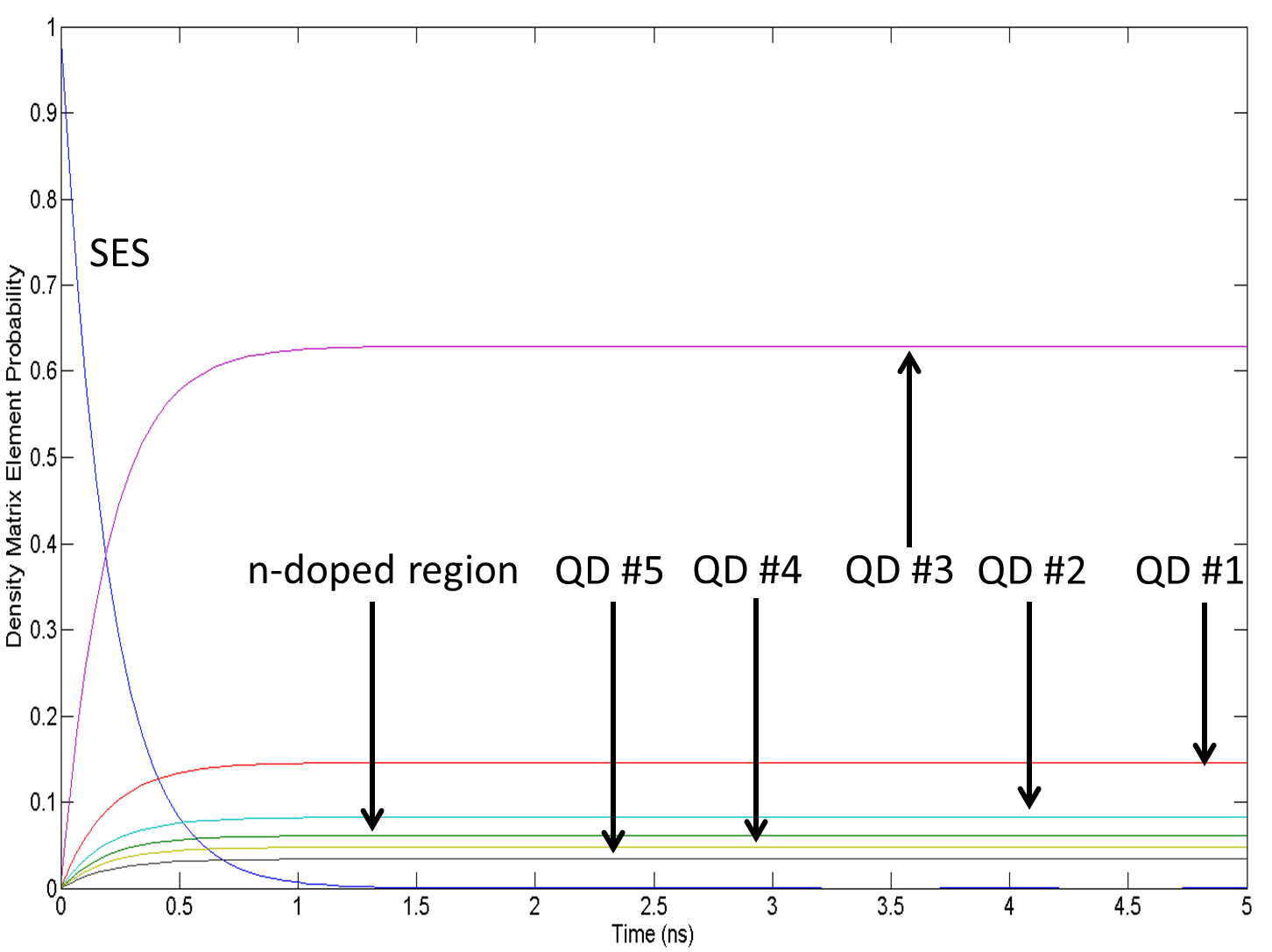}

\caption{The electron's time-dependent probability distribution among the five quantum dots at 200 K. The probability to trap the electron in QD \#3 is 63\%.
\label{fig:probabilities_decoherence_200K} }

\end{figure}

\begin{figure}[htb]
\includegraphics[width=8.5cm]{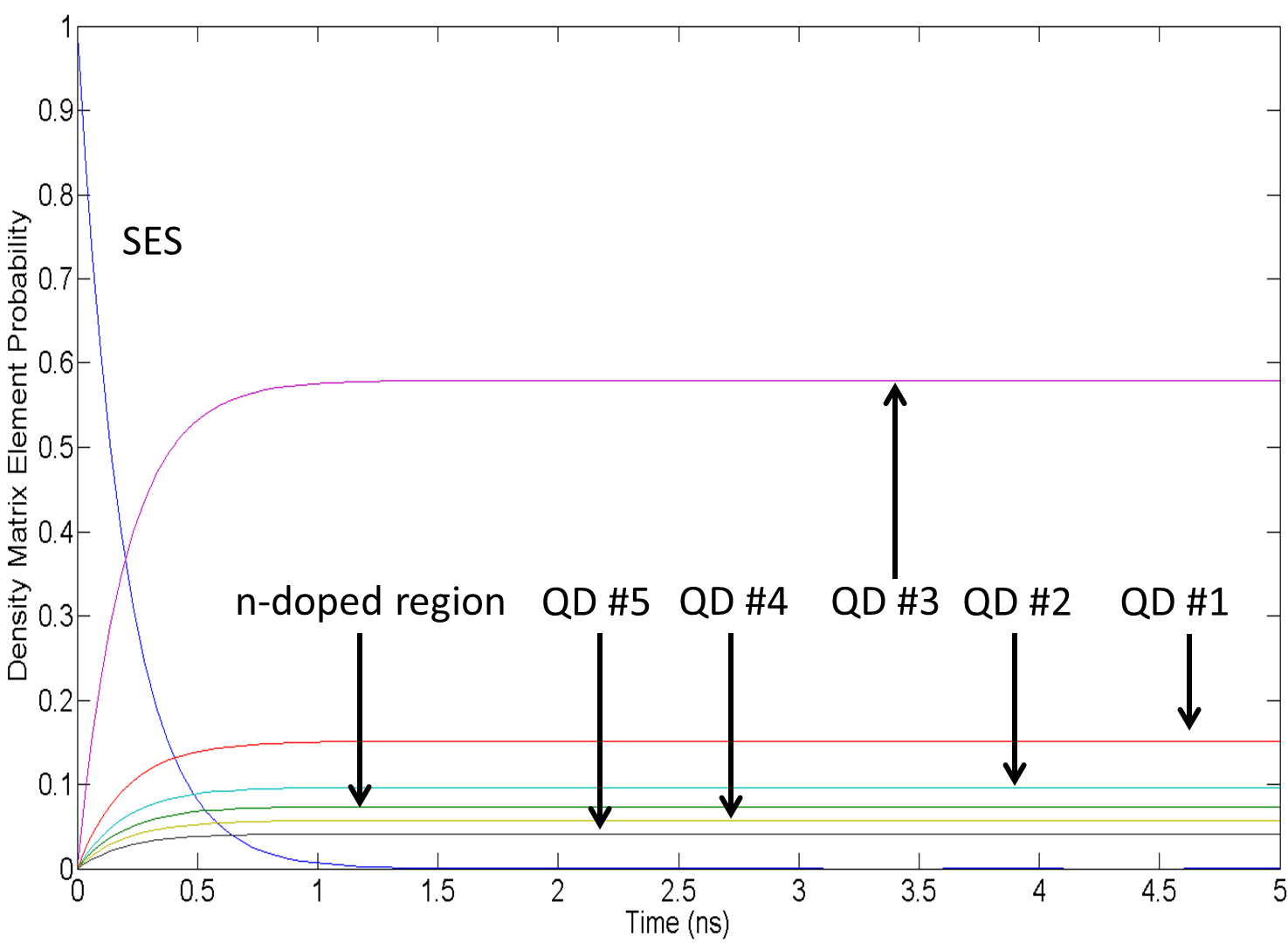}

\caption{The electron's time-dependent probability distribution among the five quantum dots at 300 K. The probability to trap the electron in QD \#3 is 58\%.
\label{fig:probabilities_decoherence_300K} }

\end{figure}

In the numerical calculations based on  Eq.~(\ref{eq:Master}) the trace of the density matrix is equal to one at all times. This ensures probability conservation, which agrees with the Hamiltonian being hermitian. In Fig.~\ref{fig:probabilities_no_decoherence}, \ref{fig:probabilities_decoherence_100K}, \ref{fig:probabilities_decoherence_200K}, and \ref{fig:probabilities_decoherence_300K} at $t=0$, ${{\rho }_{SES}}=1$, which means that there is no electron initially in the NR configuration. In addition, if decoherence is absent, the probability of the electron to get trapped in the central QD is 27\%, as shown in Fig.~\ref{fig:probabilities_no_decoherence}. In contrast, if decoherence is taken into account, the electron's trapping probability increases to 58\% - 70\% depending on the temperature (see Figs.~\ref{fig:probabilities_decoherence_100K}, \ref{fig:probabilities_decoherence_200K}, and \ref{fig:probabilities_decoherence_300K}). The probability of the electron's trapping at different temperatures and their corresponding dephasing times are shown in Table \ref{tab:trapping}.

\begin{table}[h]
\begin{center}
  \begin{tabular}{| c | c | c | }
    \hline
    Temperature [K] & Dephasing time $T_2$ [21] & Electron trap [\%] \\ \hline
    100 & 2 ps & 70 \\ \hline
    150 & 667 fs & 65 \\ \hline
    200 & 500 fs & 63 \\ \hline
    250 & 334 fs & 60 \\ \hline
    300 & 285 fs & 58 \\
    \hline
  \end{tabular}
\end{center}
\caption{Probability of electron being trapped in the central quantum dot at various temperatures between 100 K and 300 K.}
\label{tab:trapping}
\end{table}

\begin{figure}[htb]
\includegraphics[width=8.5cm]{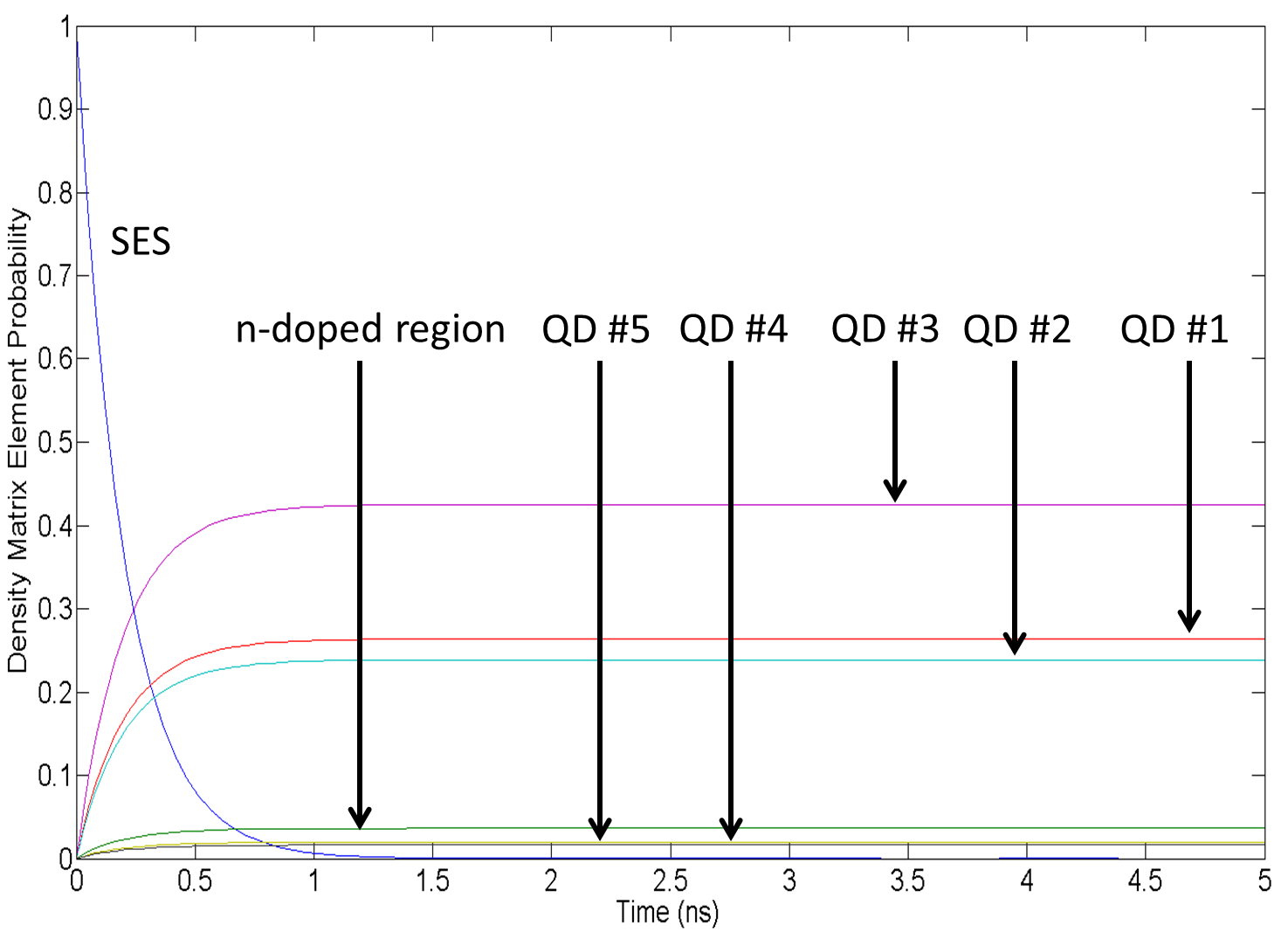}

\caption{The electron's time-dependent probability distribution among the five quantum dots, at T = 300 K  \textit{without electron pocket}. \label{fig:probabilities_decoherence_noelectronpocket} }

\end{figure}

Although the trapping probability decreases with increasing dephasing rate, it is larger than the trapping probability in the case of  vanishing decoherence. Many factors contribute to this counter-intuitive result. The first factor is the \textit{electron pocket}, which is essential for the electron to get accumulated in the central QD. To show the importance of the \textit{electron pocket}, consider the following configuration where there is no \textit{electron pocket} (i.e $t_{42} < t_{43}$ and $t_{32}$). The electron is delocalized among all five QDs. In Fig. \ref{fig:probabilities_decoherence_noelectronpocket}, the same eigenenergies and the same hopping matrix elements values were considered except for $t_{42}$. The result is due to the detuning of the energy levels of the QDs. However, there is no significant localization of the electron in QD \#3.

The second factor is the fast electron transition from the "zero-region" to QD \#1 and from QD \#5 to the "zero-region". The electron's transition rate from the "zero-region" to the first QD's ground state ${{W}_{0\to 1g}}$ is almost 100 times larger than ${{W}_{1g\to 0}}$. The electron's transition rate from the ground state of QD \#5 to the "zero-region" ${{W}_{5g \to 0}}$ is almost 10 times larger than ${{W}_{0 \to 5g}}$, which is due the n-doping of QD \#5. In this work, ${{W}_{0\to 1g}}=9.5~\times ~{{10}^{13}}~{{s}^{-1}}$ and ${{W}_{5g\to 0}}=2.0~\times ~{{10}^{14}}~{{s}^{-1}}$. These fast transitions are achieved through two factors: the DOS of the NR and the n-doping of the "zero-region" and QD \#5. Another contributing factor is the QDs' eigenenergies relative to each other, i.e. if any of the QD's energy level is modified, without adjusting the other QDs' eigenenergies, the trapping efficiency will decrease. Furthermore, the central QD's eigenenergy is the second lowest among all QDs. Based on the geometry and the dimensions of the QDs, the energy difference between the central QD's eigenenergy and the neighbor QDs' eigenenergies is almost 65 meV. This means the trapped electron needs to absorb two LO phonons in addition to a LA phonon to be able to escape. In order for the trapped electron to escape to QD \#5 ground state, an energy difference of 86 meV needs to be overcome by the emission of at least two LO phonons and one LA phonon. Even with strong ELOPI, these processes take more than 1 ns.\cite{Inoshita:1992}

Once the electron gets injected from the "zero-region" to the QD\#1, it keeps hopping among the five QDs. As shown in Sec. IV, both $t_{21}$ and $t_{54}$ are of the order of 170 meV, which means the electron hopping is faster than the phonons' response. However, due to the presence of the \textit{electron pocket}, i.e $t_{42} > t_{43} + t_{32}$, the electron hops faster between QD \#2 and QD \#4 than hopping between QD \#2 and QD \#3. Therefore electron trapping due to detuning is excluded. Meanwhile, due to the realtively smaller value of $t_{43}$ and $t_{32}$, part of electron's wavefunction slowly keeps accumulating inside QD \#3, while the rest of the electron's wavefunction hops to QD \#4, from QD \#2 then quickly to QD \#5, and then to the "zero-region", from which it hops back to QD \#1. The revolved part of the electron's wavefunction interferes constructively with the part that remained in the electron-pocket, leading to the localization of the electron. However, the localized electron will not stay inside QD \#3 forever because both $t_{43}$ and $t_{32}$ are not zero. If there is no decoherence, the electron will hop out from QD \#3 leaving behind 27$\%$ probability of trapping. The trapping probability depends on when the decoherence will terminate the hopping, i.e off-diagonal terms, between QDs. At 100 K, the electron gets accumulated inside QD \#3, before the electron starts to hop out of QD \#3, the decoherence destroys the electron's hopping, hence the electron is trapped with a trapping probability of 70$\%$. At 300 K, decoherence is much faster than at 100 K. While the electron is accumulated inside QD \#3, the decoherence inhibits further accumulation. Thus the trapping efficiency decreases to 58$\%$. The whole mechanism is based on a configuration that localizes the electron inside QD \#3 through constructive quantum interference of the electron with itself and by means of decoherence that prevents the electron from hopping out. Therefore, trapping an electron inside QD \#3 requires a delicate balance between how fast the electron is accumulated versus how large the decoherence rate is. Such a physical mechanism should manifest itself through oscillations in the diagonal density matrix elements in Fig. \ref{fig:probabilities_decoherence_different_initial_conditions}. The absence of oscillations due to the above explanation is attributed to the strong damping of the decoherence. This means the time evolution is dominated by overdamping. However, compare this results with the result obtained in App. \ref{app:Different_initial_conditions}, when the electron's initial state is $\left|1_{g},0\right\rangle$ instead of $\left|SES\right\rangle$. In that case the oscillations reflecting the electron's dynamics are visible and clearly as shown in Fig.~\ref{fig:probabilities_decoherence_different_initial_conditions}.  

\section{Conclusion}

We propose a realistic configuration to trap an electron at high temperature (100 K - 300 K) by taking advantage of the interplay between quantum interference and decoherence in an electron-pocket configuration. We would like to emphasize that the electron trapping takes place as a result of the interplay of both decoherence and quantum interference. Neither quantum interference nor decoherence alone can achieve electron trapping. In addition, as mentioned previously, the electron will require two LO phonons in addition to one LA phonon to be able to escape to the other QDs. The trapping is achieved with a probability $\rho_{33}$ depending on the temperature. At $T = 100$ K, 200 K, 300 K the trapping probability is $\rho_{33}=70$\%, 63\%, and 58\%, respectively. 

\begin{acknowledgments}
We acknowledge support from NSF (Grant No. ECCS-0725514), DARPA/MTO
(Grant No. HR0011-08-1-0059), NSF (Grant No. ECCS-0901784), AFOSR
(Grant No. FA9550-09-1-0450), and NSF (Grant No. ECCS-1128597). We thank Winston Schoenfeld, Volodymyr Turkowski, and Mikhail Erementchouk for useful
discussions.
\end{acknowledgments}

\appendix

\section{Electron-Phonon Interaction in Quantum Dots}
The Fr\"ohlich Hamiltonian describes the electron-LO phonons interaction (ELOPI). For quantum dots (QDs), it is given by
\begin{equation}
H_{e-ph} = \frac {1}{\sqrt V} \sum_{\mathbf{q},i,j} M_{\mathbf{q},i,j}a_{i}^\dagger a_{j} \left ( b_{\mathbf{q}}^\dagger + b_{\mathbf{-q}} \right) ,
\end{equation}
\begin{eqnarray}
M_{\mathbf{q},i,j} & = & i \sqrt{4 \alpha} \frac{\hbar \omega_{\mathbf{q}}}{q} \left(\frac{\hbar}{2m^* \omega_{\mathbf{q}}}\right)^{1/4} \left\langle i \rm{|} e^{i\mathbf{q}\cdot \mathbf{r}} \rm{|}  j\right\rangle, \nonumber\\
\end{eqnarray}
where $\alpha$ is the Fr\"ohlich coupling constant. In bulk InAs $\alpha=0.052$. However, in InAs QDs $\alpha=0.15$.  \cite{Heitz:1999,Odnoblyudov:1999} $\left|i\right>$ is the wavefunction for the electron level $i$ in the QD. $V$ is the volume of the NR. The phonons are assumed to be the same as those in the bulk In$_{0.45}$Ga$_{0.55}$As, since 85\% of the NR is made of In$_{0.45}$Ga$_{0.55}$As. In this work, the LO phonons are regarded as dispersionless, i.e. $\hbar \omega_{\mathbf{q}} = \hbar \omega_{LO} = 32$ meV.
We calculate the strength of couplings, in each QD, for the following cases: $\left|i\right> = \left|j\right> = \left|g\right>$, $\left|i\right> = \left|j\right> = \left|e\right>$, and ($\left|i\right> = \left|g\right>$ and $\left|j\right> = \left|e\right>$), where $\left|g\right>$ ($\left|e\right>$) is the ground (excited) state. The coupling strength in Eq.~(\ref{eq:Hamiltonian}) is calculated as follows:
\begin{eqnarray}
\lambda ^2 & = & \mathop \sum_{\left|\mathbf{q}\right|\le 2\pi/L} \left|\lambda_{\mathbf{q}} \right|^2 \nonumber\\
&  = & \frac{1}{V} \mathop \sum_{\left|\mathbf{q}\right|\le 2\pi/L} \left| M_{\mathbf{q},i,j}\right|^2.
\end{eqnarray}
For both $M_{\mathbf{q},g,g}$ and $M_{\mathbf{q},e,e}$, in all five QDs, $g = \lambda/\left(\hbar \omega_{LO} \right)$ = 0.066. Consequently, $ e^{-2g^2}\approx 1$. As for $M_{\mathbf{q},g,e}$ or $M_{\mathbf{q},e,g}$, in all five QDs, they are almost 0.0066.

\section{Energy levels and wavefunctions of cylindrical QDs}
As shown in Fig. \ref{fig:NRsetup}, the "intrinsic-region" constitutes 15\% of the NR. In addition, the "intrinsic-region" is along the z-axis. In this work, we consider the "intrinsic-region" curvature to be small. Thus, the "intrinsic-region" will be treated without curvature. Starting with Schr\"odinger's equation in cylindrical coordinates
\begin{eqnarray}
E \Psi \left(\rho, \phi, z \right) & = &
-\frac {\hbar^2}{2m^*} \left[\frac {1}{\rho} {\partial_\rho} \left(\rho {\partial_\rho} \Psi\left(\rho, \phi, z\right)\right) \right. \nonumber\\
& & +\left. \frac {1}{\rho^2} \partial_{\phi\phi} \Psi\left(\rho, \phi, z\right) + {\partial_{zz}} \Psi\left(\rho, \phi, z\right) \right] \nonumber\\
& &+ V \left(\rho, \phi, z \right)\Psi \left(\rho, \phi, z \right) , 
\end{eqnarray}
where $ E = E_{\rho} + E_z$ is the total eigenenergy of the electron. Applying separation of variables, the azimuthal differential equation and its normalized solution will be
\begin{equation}
{\partial_{\phi\phi}} \Phi \left( \phi \right) + m^2 \Phi \left( \phi \right)  =  0,
\end{equation}
and
\begin{equation}
\Phi \left( \phi \right)  =  \frac{1}{\sqrt {2\pi}} e^{+ i m \phi},
\end{equation}
where m is the azimuthal quantum number. As for the axial differential equation, it is as follows
\begin{equation}
-\frac {\hbar^2}{2m^*} Z \left( z \right) + V \left( z \right) = E_z Z \left( z \right).
\end{equation}
The solution for the axial equation is shown later. As for the radial differential equation, it is given by
\begin{equation}
{\partial_{\rho\rho}} R\left(\rho \right) + \frac {1}{\rho} {\partial_{\rho}} R \left(\rho \right) + \left( k_{\rho}^2 - \frac {m^2}{\rho^2} \right) R \left(\rho \right) = 0.
\end{equation}
The general solution for the radial differential equation is 
\begin{equation}
R \left(\rho \right) = C_{1} J_m \left( k_{\rho} \rho \right) + C_{2} N_m \left( k_{\rho} \rho \right),
\label{Radialgeneral}
\end{equation}
where $J_{m}$ is the Bessel function of the first kind and $N_{m}$ is the Bessel function of the second kind. In this work, not all QDs share the same boundary conditions, i.e \textit{electron pocket}. Both QD \#1 and \#5 do not share an interface with the \textit{electron pocket}, Thus at $\rho = 0$, $R \left( \rho \right)$ is finite, i.e $C_{2} = 0$. For both QD \#1 and \#5, at $\rho = r$, $R\left( \rho \right)$ is equal to zero. In order to satisfy this boundary condition, $k_{\rho} = \frac {\alpha_{mn}}{r} $. The radial wavefunction, for both QD \#1 and \#5, is
\begin{equation}
R \left(\rho \right) =  C_{1} J_{m} \left( \frac {\alpha_{mn} \rho}{r} \right).
\end{equation}
The energy of the electron inside either QD \#1 or QD \#5 is the sum of the radial energy and the axial energy. 
\begin{equation}
E = E_{\rho} + E_z.
\end{equation}
The axial energy will be explored in details in App. C. As for the radial energy, it is as follows
\begin{equation}
E_{\rho} = \frac {\hbar^2 \alpha_{mn}^2}{2 m^* r^2}.
\end{equation} 
$C_1$ is determined from the normalization condition as follows
\begin{eqnarray}
|C_{1}|^2 \int \limits_0^{\rho = r} \rho \left| J_{m} \left( \frac {\alpha_{mn} \rho}{r} \right) \right|^2 d\rho = 1.
\end{eqnarray}
As for QD \#2 and \#4, both are identical. 

\begin{figure}[htb]
\includegraphics[width=8.5cm]{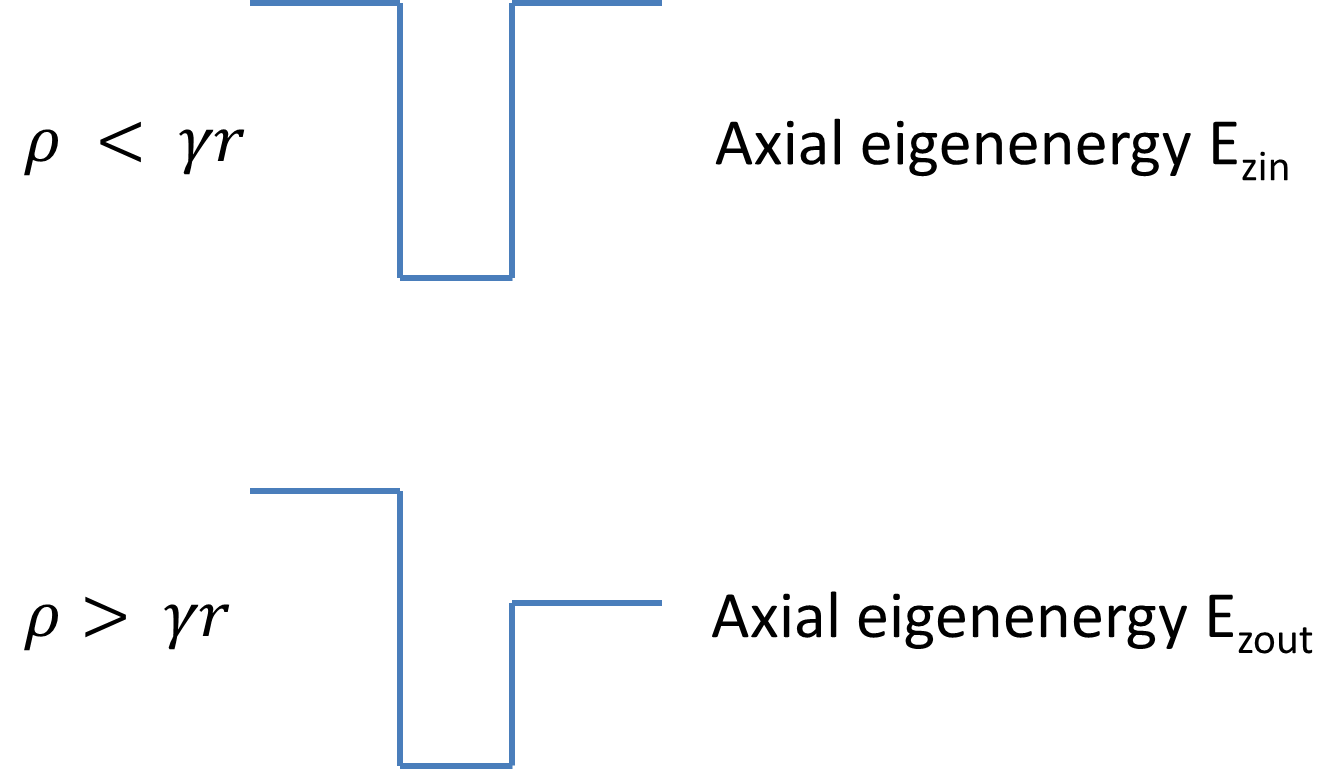}

\caption{Schematic for the QD \#2 or \#4 interface with the electron pocket.}

\label{fig:electron_pocket}
\end{figure}

For $\rho < \gamma r$, the electron's energy is denoted by $E_{in}$, where $E_{in} = E_{\rho in} + E_{zin}$, while for $\rho > \gamma r$, the electron's energy is denoted by $E_{out}$, where $E_{out} = E_{\rho out} + E_{zout}$ . Based on Fig.~\ref{fig:electron_pocket}, the boundary conditions are as follows:
\begin{enumerate}
\item at $\rho = 0$, $R \left( \rho \right)$ is finite.
\item at $\rho = \gamma r$,  $R_{in} \left(\rho \right) = R_{out} \left(\rho \right)$. 
\item at $\rho = \gamma r$, $\frac{dR_{in}\left( \rho \right)}{d\rho} = \frac{dR_{in}\left( \rho \right)}{d\rho} $.
\item at $\rho = r$, $R_{out}\left( \rho \right) = 0$. 
\item Normalization condition: $\left|\left<R|R\right>\right|^2=1$.
\item $E_{in} = E_{out}$.
\end{enumerate}
$\gamma$ is a constant factor that varies between 0 and 1. We set $\gamma$ to be 0.533. The choice for $\gamma$ is based on two factors. First, it is chosen to increase the efficiency of trapping by making $t_{42}$ larger than $t_{32} + t_{43}$. Second, the value of $\gamma$ makes the energy separation between the ground state's energy of QD \#2 or \#4 almost 70 meV higher than the ground state's energy of QD \#3. Hence it takes the trapped electron almost 1 ns to escape. For $\rho < \gamma r$, the radial differential equation is 
\begin{equation}
{\partial_{\rho\rho}} R\left(\rho \right) + \frac {1}{\rho} {\partial_{\rho}} R \left(\rho \right) + \left( k_{\rho in}^2 - \frac {m^2}{\rho^2} \right) R \left(\rho \right) = 0,
\label{eq:DiffEq.in}
\end{equation}
where $E_{\rho in} = \frac {\hbar^2 k_{\rho in}^2}{2 m^*}$. The general solution for Eq.~(\ref{eq:DiffEq.in}), is 
\begin{equation}
R_{in} \left(\rho \right) = C_{3} J_m \left( k_{\rho in} \rho \right) + C_{4} N_m \left( k_{\rho in} \rho \right).
\end{equation}
From boundary condition 1, $C_{4}$ is zero.  For $\rho > \gamma r$, the radial differential equation is 
\begin{equation}
{\partial_{\rho\rho}} R\left(\rho \right) + \frac {1}{\rho} {\partial_{\rho}} R \left(\rho \right) + \left( k_{\rho out}^2 - \frac {m^2}{\rho^2} \right) R \left(\rho \right) = 0,
\label{eq:DiffEq.out}
\end{equation}
where $E_{\rho out} = \frac {\hbar^2 k_{\rho out}^2}{2 m^*}$. The general solution for Eq.~(\ref{eq:DiffEq.out}), is 
\begin{equation}
R_{out} \left(\rho \right) = C_{5} J_m \left( k_{\rho out} \rho \right) + C_{6} N_m \left( k_{\rho out} \rho \right).
\end{equation} 
Both $R_{in}$ and $R_{out}$ share the same "m". From boundary condition 4, 
\begin{equation}
C_6 = -C_{5} \frac {J_m \left(k_{ \rho out} r\right)}{N_m \left(k_{ \rho out} r\right)}.
\end{equation}
From both boundary conditions 2 and 3, and simple algebraic manipulation, the transcendental equation reads
\begin{eqnarray}
\lefteqn{
k_{\rho out} J_{m}\left( k_{\rho in} \gamma r \right) \left[ J_m ^{'} \left(k_{\rho out} \gamma r\right) N_m \left(k_{\rho out} \gamma r\right)\right. 
} \nonumber\\
\lefteqn{-\left. J_m \left(k_{\rho out} \gamma r\right) N_m ^{'} \left(k_{\rho out} \gamma r\right)\right]  
}\nonumber\\
& = & k_{\rho in} J_{m}^{'}\left( k_{\rho in} \gamma r \right) \left[ J_m \left(k_{\rho out} \gamma r\right) N_m \left(k_{\rho out} r\right) \right. \nonumber\\
& & -\left. J_m \left(k_{\rho out} r\right) N_m \left(k_{\rho out} \gamma r\right)\right]. 
\end{eqnarray}

\begin{figure}[htb]
\includegraphics[width=8.5cm]{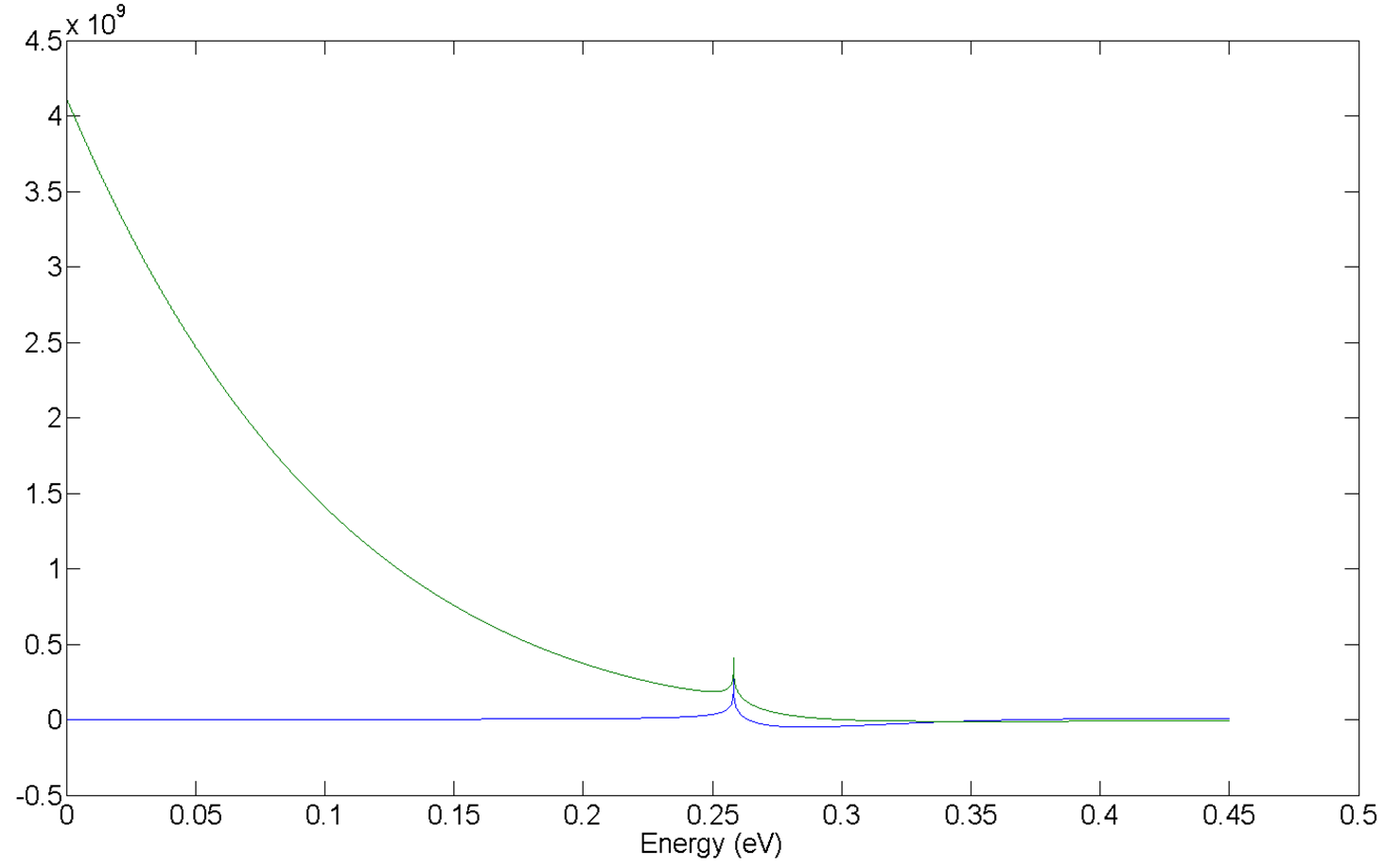}

\caption{The solution for the transcendental equation.}
\label{fig:transcendental}

\end{figure}

As shown in Fig.~\ref{fig:transcendental}, the solution for this transcendental equation, with $\gamma = 0.533$, is $E = 0.348$ eV. QD \#3 is treated the same as both QD \#1 and \#5, but with different radius.

\section{The axial energy levels of the cylindrical QDs}

\begin{figure}[htb]
\includegraphics[width=4.0cm]{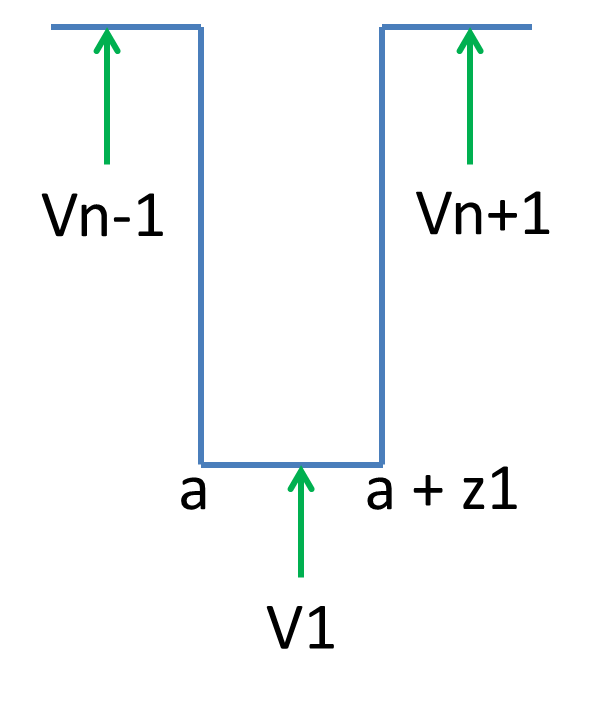}

\caption{Schematic for the QD's height.}

\end{figure}

We write down the solution for Schr\"odinger's non-relativistic time-independent equation for the QD's different regions along the z-axis,
for $-\infty < z < a$,
\begin{equation}
Z_1 \left(z \right) = A e^{{k_{n-1}} \left( z - a \right)},
\label{eq:A1}
\end{equation}
for $ a < z < a + z_1$,
\begin{equation}
Z_2 \left(z \right) = B \cos\left(k_n \left( z - a \right)\right) + C \sin\left(k_n \left( z - a \right)\right),
\label{eq:A2}
\end{equation}
where $z_1$ is the height of the QD. For $ a + z_1 < z < \infty$,
\begin{equation}
Z_1 \left(z \right) = D e^{-{k_{n+1}} \left( z - a \right)}.
\label{eq:A3}
\end{equation}
From the boundary conditions at z = a, we obtain
\begin{eqnarray}
B & = & A, \\
\label{eq:A4}
C & = &  \sigma_1 A,
\label{eq:A5}
\end{eqnarray}
where $\sigma_1 = \frac {m_n k_{n-1}}{m_{n-1} k_n}$. From the boundary conditions at $z = a + z_1$, we get
\begin{equation}
B \cos\left( k_n z_1 \right) + C \sin\left( k_n z_1 \right) = D e^{-{k_{n+1}} \left( z_1 \right)}
\label{eq:A6}
\end{equation}
and 
\begin{equation}
-\frac {k_n}{m_n} B \sin\left( k_n z_1 \right) + \frac {k_n}{m_n} C \cos\left( k_n z_1 \right) = -\frac {k_{n+1}}{m_{n+1}} D e^{-{k_{n+1}} \left( z_1 \right)}.
\label{eq:A7}
\end{equation}

From Eq.~(\ref{eq:A1}) and Eq.~(\ref{eq:A2}), Eq.~(\ref{eq:A3}) and Eq.~(\ref{eq:A4})  yield, respectively,

\begin{equation}
A \cos\left( k_n z_1 \right) + \sigma_1 A \sin\left( k_n z_1 \right) = D e^{-{k_{n+1}} \left( z_1 \right)},
\label{eq:A8}
\end{equation}
and 
\begin{equation}
-A \sin\left( k_n z_1 \right) + \sigma_1 A \cos\left( k_n z_1 \right) = -\sigma_2 D e^{-{k_{n+1}} \left( z_1 \right)},
\label{eq:A9}
\end{equation}
where $\sigma_2 = \frac {m_n k_{n+1}}{m_{n+1} k_n}$. After getting rid of $A$ by dividing Eq.~(\ref{eq:A8}) by Eq.~(\ref{eq:A9}) and using a few straightforward algebraic steps, the transcendental equation for the QD eigenenergies is 

\begin{equation}
 \tan \left(k_n  z_1 \right) = \frac {\sigma_2 + \sigma_1}{1 - \sigma_1 \sigma_2}.
\label{eq:A10}
\end{equation}
A graphical solution provides the eigenenergies for each QD.

%PROVIDE THE EIGENENERGIES FOR EACH QD

Going back to Eq.~(\ref{eq:A8}) and Eq.~(\ref{eq:A9}), D in terms of A is given by
\begin{equation}
D = \frac {\sigma_1^2 + 1}{\sigma_1 + \sigma_2} \sin \left( k_n z_1 \right) e^{k_{n+1} \left( z_1 \right)} A.
\label{eq:11}
\end{equation}

After normalization, we obtain $A=1/\sqrt{N}$, where
\begin{eqnarray}
N & = & \frac {1}{2 k_{n-1}} + \left( \frac {\sigma_1^2 + 1}{\sigma_1 + \sigma_2} \right)^2 \sin^2 \left( k_n z_1 \right) \frac {1}{2 k_{n+1}} 
\nonumber\\
& & + \frac{1}{4 k_n} \left( 2 k_n z_1 + \sin \left( 2 k_n z_1 \right) \right) 
\nonumber\\
& & + \sigma_1^2 \left( \frac {z_1}{2} - \frac {\sin \left( 2 k_n z_1 \right)}{4 k_n}\right) \nonumber\\
&& - \frac {\sigma_1}{2 k_n} \left( \cos \left( 2 k_n z_1 \right) - 1 \right).
\label{eq:A12}
\end{eqnarray}

\section{Electron's oscillations among the five quantum dots}
\label{app:Different_initial_conditions}
\begin{figure}[htb]
\includegraphics[width=8.5cm]{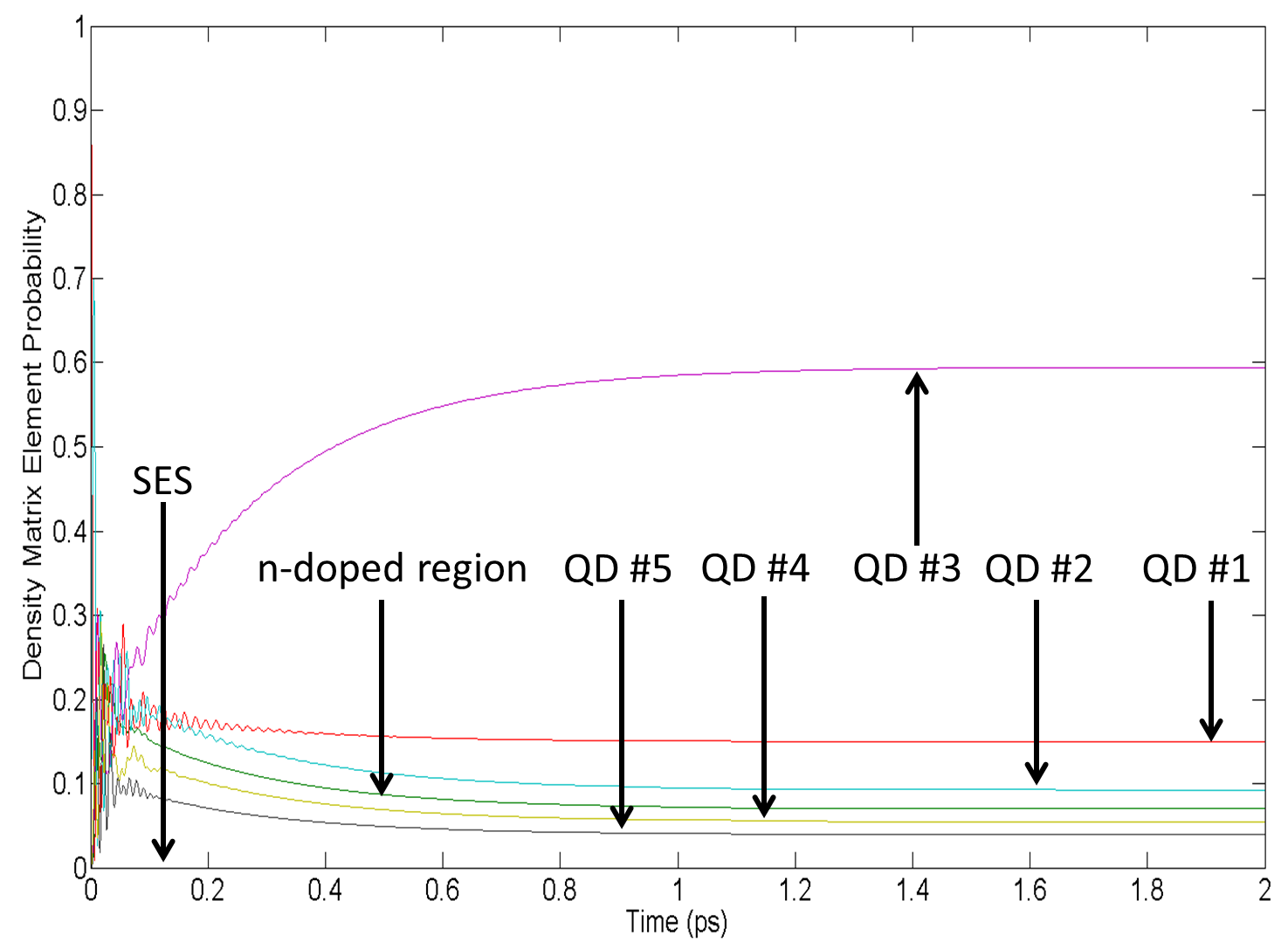}

\caption{The electron's time-dependent probability distribution among the five quantum dots at room temperature with $\rho_{11}$  = 1 as initial condition.}

\label{fig:probabilities_decoherence_different_initial_conditions}
\end{figure}

In this appendix we change the initial condition from $\rho_{SES}$ = 1 to $\rho_{11}$ = 1. The reason is to illustrate the oscillations that are suppressed by decoherence if the initial condition is $\rho_{SES}$ = 1. In Fig.~\ref{fig:probabilities_decoherence_different_initial_conditions} oscillations due to interference are clearly visible. Note that despite the change in the initial condition, the results are the same. In addition, the time scale here is in the picosecond regime, while the timescale for the previous results is in the nanosecond regime. The reason is that the injection rate from the SES to the NR is $5~\times ~{{10}^{9}}~{{s}^{-1}}$.

\section{Approximation of the hopping matrix elements}
\label{app:Landau}
\begin{figure}[htb]
\includegraphics[width=8.5cm]{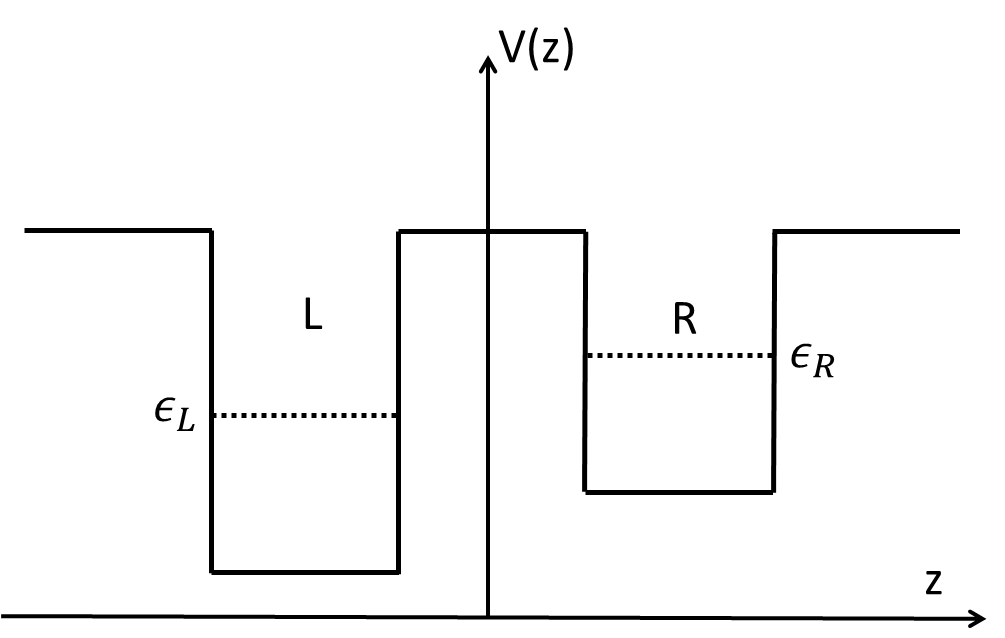}

\caption{Schematic showing an asymmetric double well potential.}
\label{fig:double_well}
\end{figure}

As an illustration, in this appendix we are going to generalize the 1D calculation of the hopping matrix elements for a symmetric double-well potential presented in Ref.~\onlinecite{Landau&Lifshitz} to the case of an asymmetric double-well potential.
We emphasize that we do not use this approximation in the calculations. This appendix is provided for educational purpose only.
The effective Hamiltonian for a coupled two-level system is
\begin{equation}
H=\left(\begin{array}{cc}
\epsilon & t \\
t & -\epsilon
\end{array}\right),
\end{equation}
where $2\epsilon$ is the bias between the right and left well, and $t$ is the hopping matrix element.
The eigenenergies are $E_\pm=\sqrt{\epsilon^2+t^2}$. The corresponding eigenstates are
\begin{eqnarray}
\left|\psi_+\right> & = & \cos\frac{\theta}{2}\left|\psi_R\right>+\sin\frac{\theta}{2}\left|\psi_L\right>, \\
\left|\psi_-\right> & = & -\sin\frac{\theta}{2}\left|\psi_R\right>+\cos\frac{\theta}{2}\left|\psi_L\right>,
\end{eqnarray}
where $\tan\theta=t/\epsilon$ with $0\le\theta<\pi$.
For determining $E_+$ we use the following two Schr\"odinger equations:
\begin{eqnarray}
\psi_R''+\frac{2m}{\hbar^2}\left(E_R-V\right)\psi_R & = & 0, \label{eq:SE1}\\
\psi_+''+\frac{2m}{\hbar^2}\left(E_+-V\right)\psi_R & = & 0, \label{eq:SE2}
\end{eqnarray}
where $V$ is the 1D potential shown in Fig.~\ref{fig:double_well}.
Multiplying $\psi_+$ to Eq.~(\ref{eq:SE1}), multiplying $\psi_R$ to Eq.~(\ref{eq:SE2}), and taking the difference results in
\begin{equation}
\psi_+\psi_R''-\psi_R\psi_+''+\frac{2m}{\hbar^2}\left(E_R-E_+\right)\psi_R\psi_+=0
\end{equation}
After integration from 0 to infinity and integration by parts, we obtain
\begin{equation}
E_+-E_R = \frac{\hbar^2}{2m\delta_{R,+}^+} \left[-\psi_+(0)\psi_R'(0) +\psi_R(0)\psi_+'(0)\right],
\end{equation}
where $\delta_{R,+}^+=\int_0^\infty\psi_R\psi_+ dx$.
%$\psi_+$ can be evaluated as follows:
%\begin{equation}
%\psi_+(0)=\left[\delta_{+,R}\psi_R(0)+\delta_{+,L}\psi_L(0)\right],
%\end{equation}
%where $\delta_{+,R}=\int_{-\infty}^\infty\psi_+^*\psi_R dx$ and $\delta_{+,L}=\int_{-\infty}^\infty\psi_+^*\psi_L dx$.

%

A similar calculation can be done for determining $E_--E_R$. Evaluating the difference, we obtain
\begin{align}
E_--E_+ = & \frac{\hbar^2}{2m\delta_{R,-}^+} \left[-\psi_+(0)\psi_L'(0) +\psi_L(0)\psi_+'(0)\right] \nonumber\\
& -\frac{\hbar^2}{2m\delta_{R,+}^+} \left[-\psi_+(0)\psi_R'(0) +\psi_R(0)\psi_+'(0)\right] ,
\end{align}
where  $\delta_{L,+}^+=\int_0^\infty\psi_L\psi_+ dx$.
Using the approximations $\delta_{R,-}^+\approx -\sin\frac{\theta}{2}$ and $\delta_{R,+}^+\approx \cos\frac{\theta}{2}$, we can substantially simplify the above equation to
\begin{eqnarray}
E_--E_+ & = & \frac{\hbar^2}{2m} \left[\cot\frac{\theta}{2}+\tan\frac{\theta}{2}\right] \nonumber\\
& & \times \left[\psi_L(0)\psi_R'(0) -\psi_L'(0)\psi_R(0)\right].
\end{eqnarray}
Using the formulas $\tan\frac{\theta}{2}=\frac{1-\cos\theta}{\sin\theta}$ and $\cot\frac{\theta}{2}=\frac{1+\cos\theta}{\sin\theta}$, we obtain
\begin{equation}
E_--E_+=\frac{\hbar^2k^2}{m\sin\theta}=2\sqrt{\epsilon^2+t^2},
\end{equation}
where we defined $k^2=\left[\psi_L(0)\psi_R'(0) -\psi_L'(0)\psi_R(0)\right]$.
Since $\sin\theta=t/\sqrt{\epsilon^2+t^2}$, we get
\begin{equation}
t=\frac{\hbar^2k^2}{2m}.
\end{equation}
We compared this approximation with the result obtained using Eq.~(\ref{eq:hopping}). Our calculations show that this approximation gives less than 50\% of the kinetic matrix element in Eq.~(\ref{eq:hopping}) and less than 40\% of the total matrix element in Eq.~(\ref{eq:hopping}).
This discrepancy is due to the approximations $\delta_{R,-}^+\approx -\sin\frac{\theta}{2}$ and $\delta_{R,+}^+\approx \cos\frac{\theta}{2}$,
which neglect the tails of the wavefunctions. This result illustrates also the importance of including the off-diagonal matrix elements of the potential.

%In order to obtain a better approximation within the tight-binding model, it is possible to calculate the eigenstates first numerically
%by diagonalizing the Hamiltonian $H_t$ for two neighboring QDs.
%The eigenstates are then written as
%\begin{equation}
%\psi_\pm=\delta_{\pm,R}\psi_R+\delta_{\pm,L}\psi_L,
%\end{equation}
%where $\delta_{\pm,R}=\int_{-\infty}^\infty\psi_\pm^*\psi_R dx$ and $\delta_{\pm,L}=\int_{-\infty}^\infty\psi_\pm^*\psi_L dx$
%are the overlap integrals.
%Using the formula $2\cot\theta=\cot\frac{\theta}{2}-\tan\frac{\theta}{2}$, we find
%\begin{equation}
%\frac{\delta_{+,R}}{\delta_{-,R}}+\frac{\delta_{+,L}}{\delta_{-,L}}=-2\cot\theta=-\frac{2\epsilon}{t}.
%\end{equation}
%Thus, we obtain
%\begin{equation}
%t=-\frac{2\epsilon}{\frac{\delta_{+,R}}{\delta_{-,R}}+\frac{\delta_{+,L}}{\delta_{-,L}}}.
%\label{eq:t_bias}
%\end{equation}
%This expression can lead to large numerical errors in the case of degeneracy because when $\epsilon\rightarrow 0$, then the denominator also vanishes.
%Therefore, for small bias it is better to use the formula $\frac{2}{\sin\theta}=\cot\frac{\theta}{2}+\tan\frac{\theta}{2}$.
%Then we obtain
%\begin{equation}
%-\frac{\delta_{+,R}}{\delta_{-,R}}+\frac{\delta_{+,L}}{\delta_{-,L}}=\frac{2}{\sin\theta}=\frac{2\sqrt{\epsilon^2+t^2}}{t}.
%\end{equation}
%For $\epsilon=0$ we get
%\begin{equation}
%t=-\frac{\delta_{+,R}}{\delta_{-,R}}+\frac{\delta_{+,L}}{\delta_{-,L}}.
%\label{eq:t_nobias}
%\end{equation}
%The formulas in Eqs.~(\ref{eq:t_bias}) and (\ref{eq:t_nobias}) give a better approximation of Eq.~(\ref{eq:hopping}).

%\bibliographystyle{natbib}
\bibliographystyle{apsrev}
\bibliography{decoherence_assisted_bibliography}

\end{document}